\documentclass[cits]{PoS}

\usepackage{xspace}

\newcommand{\ep}{$E_{\rm p}$\xspace}

\title{Untriggered \emph{Swift}-GRBs in \emph{Fermi}/GBM data}

\ShortTitle{Untriggered GRBs in GBM data}

\author{\speaker{David Gruber} on behalf of the \emph{Fermi}/GBM collaboration\\
          Max Planck Institute for extraterrestrial Physics, Giessenbachstr. 1., 
  \\85748 Garching, Germany\\
        E-mail: \email{dgruber@mpe.mpg.de}}

\abstract{The \emph{Fermi} Gamma-Ray Burst Monitor (GBM) onboard the Fermi spacecraft currently operates on several trigger algorithms on various time scales and energy ranges.
Motivated by the pursuit of faint Gamma-Ray Bursts (e.g. the elusive class of postulated low-luminosity GRBs), here we present the search for untriggered GRBs in the GBM data stream. To this end, I will demonstrate the methods and algorithms which have been developed by the GBM team.
As a preliminary result, I am going to highlight the spectral analysis of GRBs which triggered the Swift satellite, but not GBM, and came from positions above the horizon, with a favorable orientation to at least one GBM detector. The properties of these GRBs are then compared to the full sample of GBM GRBs published in the GBM spectral catalogue.
We estimate that the lower limit for untriggered GRBs in the GBM data is about 1.6 GRBs per month which corresponds to about 7\% of the triggered GRBs.
}

\FullConference{Gamma-Ray Bursts 2012 Conference -GRB2012,\\
		May 07-11, 2012\\
		Munich, Germany}

\begin{document}

\section{Introduction}

The Gamma-Ray Burst Monitor (GBM) is one of the instruments onboard the \emph{Fermi} Gamma-Ray Space Telescope \cite{atwood09} launched on June 11, 2008. Specifically designed for GRB studies, GBM is comprised of a total of 12 sodium iodide (NaI(Tl)) scintillation detectors covering the energy range from 8 keV to 1~MeV and two bismuth germanate scintillation detectors (BGO) sensitive to energies between 150~keV and 40~MeV \cite{meegan09}. %Thus, GBM offers a unprecedented view of GRBs, covering more then 3 decades in energy.

\begin{figure}[htbp]
\begin{center}
\includegraphics[width=0.7\textwidth]{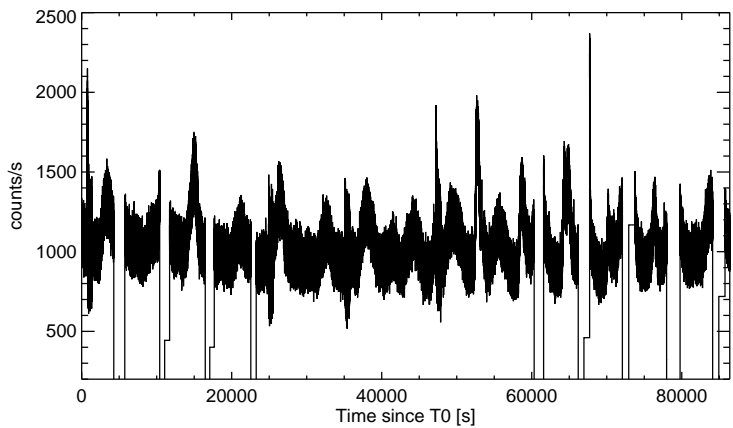}
\caption{Typical CTIME light curve of one NaI detector over a full day in the energy interval from $10-1000$~keV. One can clearly identify the overall background variations and SAA passages during which the detectors are switched-off.}
\label{fig:ctime}
\end{center}
\end{figure}

GBM continuously observes the whole unocculted sky, with its flight software (FSW) constantly monitoring the count rates recorded in the various detectors. For GBM to trigger on a GRB or any other high energy-transient, two or more detectors must have a statistically significant increase in count rate above the background rate. GBM currently operates on 75 (of 119 supported) different trigger algorithms, each defined by its timescale (from 16~ms to 4~s) and energy range ($25-50$~keV, $50-300$~keV, $>100$~keV, and $>300$~keV). In addition, the trigger algorithm is constructed to have two temporally overlapping windows (at half the window length) for all timescales above 16~ms (the so called ``offset'') and each algorithm can be operated on different threshold settings from $0.1\sigma$ to $25.5\sigma$. 
GBM persistently records two different types of science data, called CTIME (fine time resolution, coarse spectral resolution) and CSPEC (coarse time resolution, full spectral resolution). CTIME (CSPEC) data have a nominal time resolution of 0.256~s (4.096~s) (see Fig.\ref{fig:ctime}) which is increased to 64~ms (1.024~s) when GBM triggers. After 600~s in triggered mode, both data types return to the non-triggered time resolution.
%This makes triggers due to statistical fluctuations or fluorescence spikes less likely
 For a more detailed review on the data types and trigger properties of GBM we refer to \cite{meegan92} and \cite{paciesas12}.

Here, I present the preliminary results of an on-ground search in CTIME data for GRBs which were detected by the \emph{Swift} satellite \cite{gehrels04} but did not trigger GBM, although having a favorable orientation to the GBM detectors.

\section{Methods}\label{methods}

From 2008 August 22 to 2012 January 06, \emph{Swift} observed 299 GRBs\footnote{\tt{http://swift.gsfc.nasa.gov/docs/swift/archive/grb\_table.html/}}. 107 (36 \%) of these also triggered GBM. 30 (10\%) \emph{Swift}-GRBs happened while GBM was flying through the South Atlantic Anomaly (SAA) during which the high-voltage of the detectors is switched off. 88 (29\%) of the GRBs were either shadowed by the Earth when they triggered \emph{Swift} or had source angles to all detectors which were larger than $60^{\circ}$, making a GBM detection unlikely because of the rapidly dropping effective area of the NaI detectors at such high incidence angles. Using these geometrical and orientational arguments, the remaining 74 (25\%) \emph{Swift}-GRBs could have been observed by GBM. 

The newly developed on-ground trigger algorithm looks for these GRBs by employing a routine which is similar to the FSW. For a trigger, an excess count rate (above background) in CTIME data in two detectors must be observed with a statistical significance of $4\sigma$ and $3.8\sigma$, respectively. The search is restricted to the energy range between $50-300$~keV which is the energy range in which GRBs typically trigger GBM (see \cite{paciesas12}). The trigger time scales vary from 0.256~s to 8.192~s. The greatest advantage, compared to the FSW triggering, comes from the fact that the background can be handled better on-ground because we have knowledge about the background variation over the whole day which is simply modeled using a spline function. In addition to this, the rise and set times of 92 known gamma-ray sources  over the Earth's horizon with respect to GBM are calculated. Usually, these rises/sets result in an abrupt change in count rate in the NaI detectors, best described as ``occultation steps'' \cite{hodge12}. In order to account for this effect in the NaI detectors, the source's photon spectrum is used which is then folded through the detector response to calculate the expected change in count rate.
%(either the intrinsic one if it is known or approximated with $\Phi(E)\propto E^{-3}$)
\section{Results}

\begin{figure}[htbp]
\begin{center}
\includegraphics[width=0.45\textwidth]{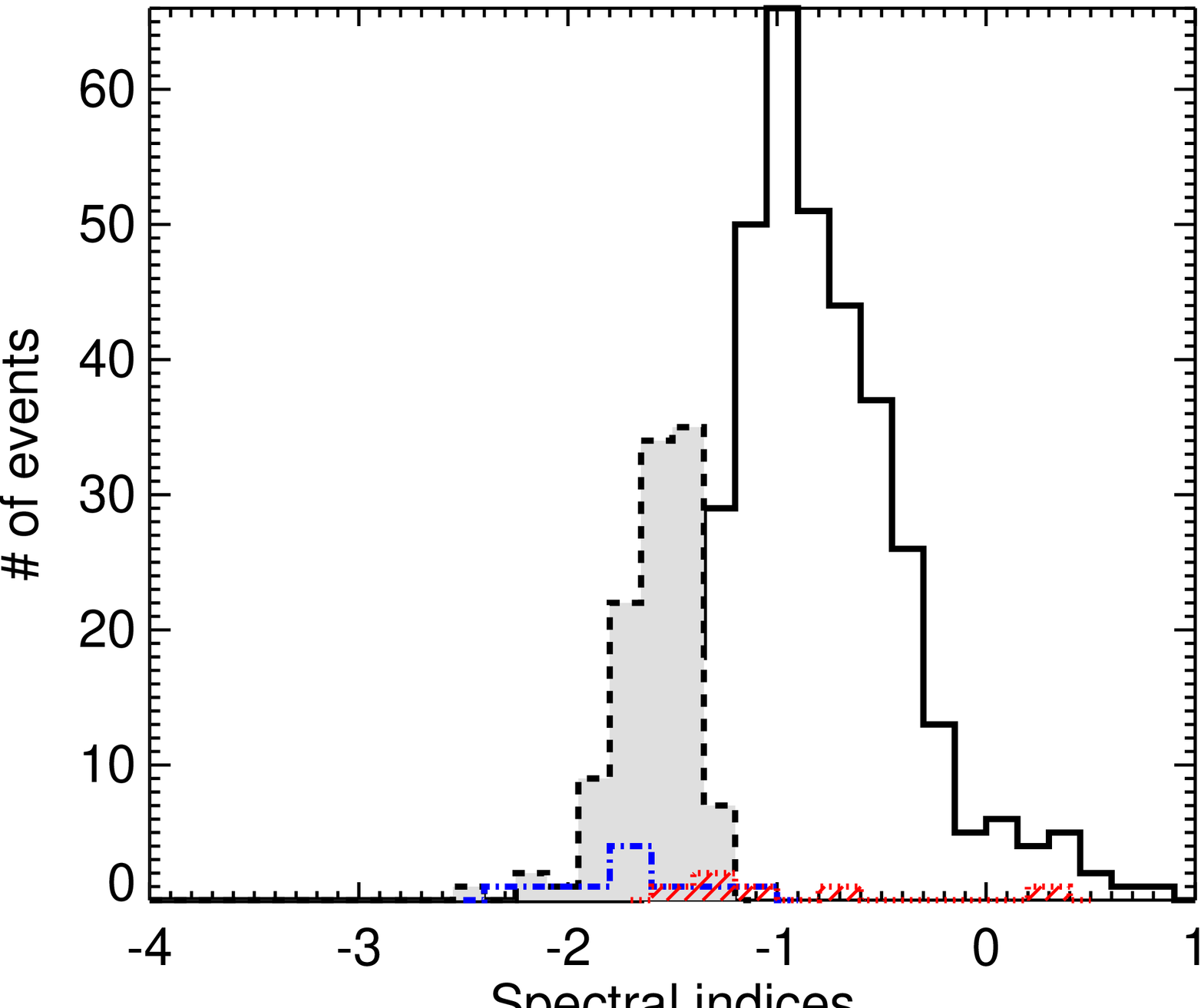}
\includegraphics[width=0.45\textwidth]{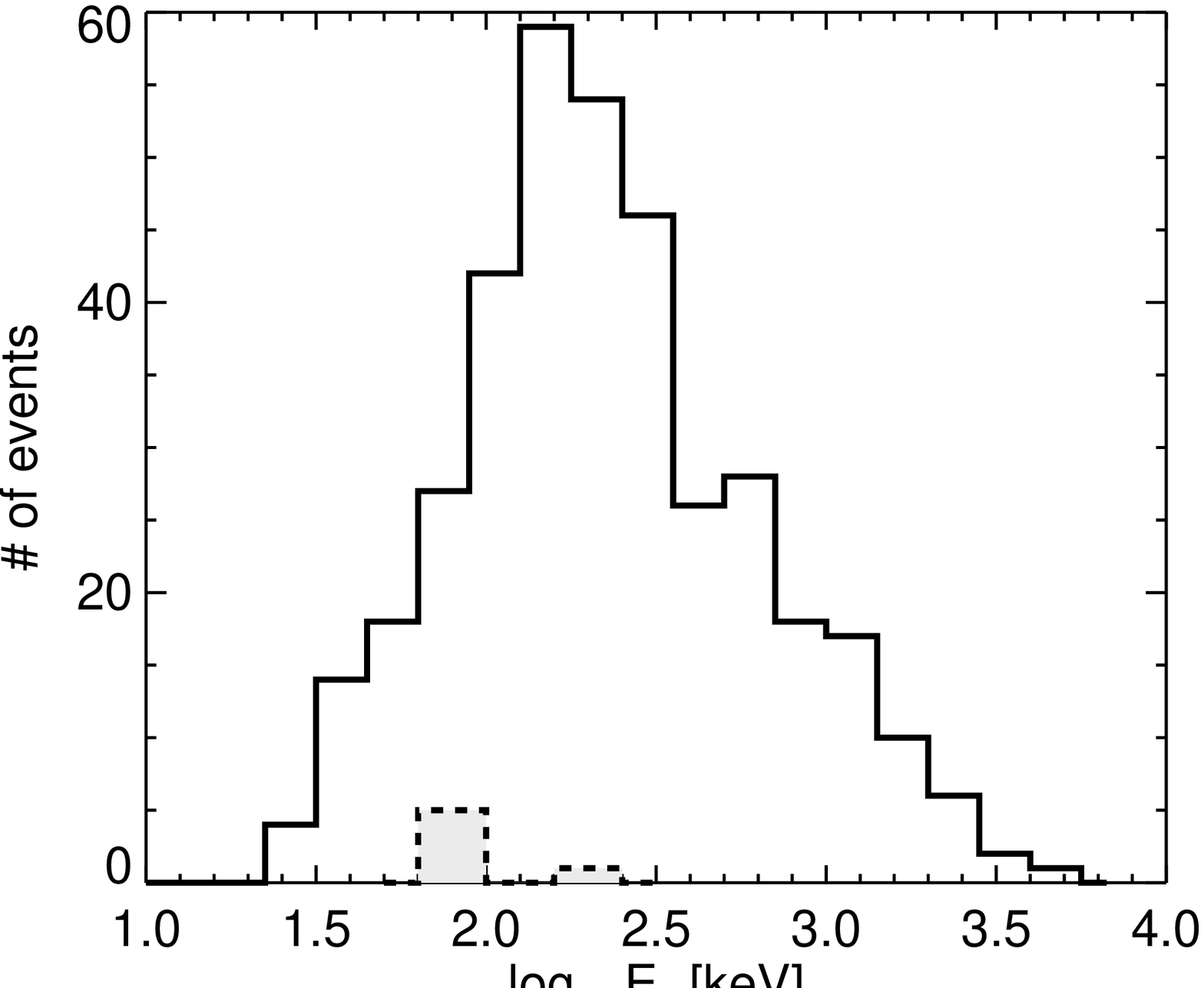}
\caption{\textit{Left panel:} Distribution of the low-energy power-law indices. The index of the PL model for the untriggered sample (blue dash-dotted histogram) and all GBM-GRBs (light-grey dashed histogram). The histogram for COMP and BAND model for the untriggered sample (red hatched histogram) and the full GBM-GRB sample (black solid histogram) are indicated. \textit{Right panel:} Distribution of the \ep values for the 6 bursts which were fitted by COMP or BAND for the untriggered sample (light-grey dashed histogram) and the full GBM-GRB sample (black solid histogram). The values for all triggered GBM-GRBs are taken from \cite{goldstein12}.}
\label{fig:specidx}
\end{center}
\end{figure}

17 untriggered GRBs (1 short and 16 long) were identified  (see Table \ref{tab:grbs}). A spectral analysis of all the long GRBs, using  CSPEC data, was performed. The model selection comprised a simple power law (PL), a power law with an exponential cutoff (COMP) and the Band function (BAND) \cite{band93a}. The spectral analysis reveals no fundamental differences or peculiarities (see Fig.\ref{fig:specidx}) compared to the full GBM-GRB sample \cite{goldstein12}. As is expected, these events are populating the lower end of the photon flux distribution with fluxes ranging between $0.6 - 1.5$~ph~cm$^{-2}$~s$^{-1}$ in the $10 - 1000$~keV energy range. Along with being faint, most of the events are also relatively soft, having \ep values which are located at the lower tail of the \ep distribution of the complete GBM-GRB sample \cite{goldstein12}. No spectral analysis could be performed for the short GRB~090305A (time resolution of CSPEC data is too long compared to the actual duration, $T_{\rm90}\sim0.3$~s,  of the GRB. However, GBM currently observes in continuous TTE mode above certain regions of the Earth. As soon as this observing mode will be extended and activated full-time, this will (i) result in more detections of untriggered short GRBs and, more importantly, (ii) allow for a spectral analysis of such short events). The remaining 57 events which could have been detected by GBM (see Section \ref{methods}) are probably to weak to be seen in GBM and are only observed by \emph{Swift} due to its better sensitivity.

Considering that the Field-of-View (FoV) of \emph{Swift} is roughly 4 times smaller than GBM's FoV, one can estimate that the GBM on-board trigger algorithm misses $\sim 1.6$ GRBs per month.
%$\frac{4\times 17\; {\rm GRBs}}{42\; {\rm months}} \sim 1.6$ 

\begin{table}
\centering
\begin{tabular}{cccccc}
\hline
GRB ID 		& model	&	E$_{\rm peak}$ [keV]		&	$\alpha$			&		$\beta$		&		flux [ph~cm$^{-2}$~s$^{-1}$]	\\
\hline
\hline
111212A 		&	PL		&	$-$						&	$-1.62 \pm 0.04$	&		$-$		&		$0.78 \pm 0.04$			\\
110801A 		&	PL		&	$-$						&	$-1.61 \pm 0.04$	&		$-$		&		$1.10 \pm 0.04$			\\
			&	PL		&	$-$						&	$-1.86 \pm 0.07$	&		$-$		&		$1.07 \pm 0.06$			\\
110719A 		&	COMP	&	$73.5\pm5.0$				&	$+0.36 \pm 0.30$	&		$-$		&		$0.63 \pm 0.04$			\\
110414A		&	BAND	&	$67.8\pm8.8$				&	$-0.66 \pm 0.21$	&  $-2.23 \pm 0.2$	&		$0.86 \pm 0.03$			\\
110411A		&	COMP	&	$64.2\pm5.3$				&	$-1.29 \pm 0.11$	&		$-$		&		$1.04 \pm 0.03$			\\
110315A 		&	COMP	&	$66.5\pm8.9$				&	$-1.58 \pm 0.10$	&		$-$		&		$1.28 \pm 0.04$			\\
110312A		&	PL		&	$-$						&	$-2.35 \pm 0.11$	&		$-$		&		$0.89 \pm 0.06$			\\
100902A 		&	COMP	&	$94.9\pm17.2$				&	$-1.31 \pm 0.14$	&		$-$		&		$0.73 \pm 0.03$				\\
100305A		&	PL		&	$-$						&	$-1.25 \pm 0.07$	&		$-$		&		$0.79 \pm 0.07$			\\
091029A		&	PL		&	$-$						&	$-1.83 \pm 0.04$	&		$-$		&		$1.51 \pm 0.05$			 \\
090728A		&	PL		&	$-$						&	$-1.60 \pm 0.05$	&		$-$		&		$1.12 \pm 0.06$			 \\
090520A		&	PL		&	$-$						&	$-1.14 \pm 0.08$	&		$-$		&		$0.74 \pm 0.08$			\\
090404A 		&	PL		&	$-$						&	$-2.20 \pm 0.05$	&		$-$		&		$1.31 \pm 0.05$			\\
090305AS 	&	$-$		&	$-$						&	$-$				&		$-$		&		$-$						\\ 
090123A		&	PL		&	$-$						&	$-1.68 \pm 0.04$	&		$-$		&		$0.73 \pm 0.03$			 \\
080906A 		&	COMP	&	$244.0\pm48.0$			&	$-1.19 \pm 0.10$	&		$-$		&		$0.92 \pm 0.04$			\\
080916B 		&	PL		&	$-$						&	$-1.55 \pm 0.12$	&		$-$		&		$0.95 \pm 0.15$			\\
\hline
\end{tabular}
\caption{\emph{Swift}-GRBs found in untriggered GBM data. The results of the spectral analysis, peak energy \ep, low-energy power law index $\alpha$, high-energy power law index $\beta$ and photon flux, are shown. Please note that GRB~110801A was a very long GRB with two distinctive emission episodes and that a spectral fit was not possible for short GRB~090305A. } \label{tab:grbs}
\end{table}

\end{document}